\documentclass[pdflatex,sn-mathphys-num]{sn-jnl}

\usepackage{graphicx}%
\usepackage{multirow}%
\usepackage{amsmath,amssymb,amsfonts}%
\usepackage{amsthm}%
\usepackage{mathrsfs}%
\usepackage[title]{appendix}%
\usepackage{xcolor}%
\usepackage{textcomp}%
\usepackage{manyfoot}%
\usepackage{booktabs}%
\usepackage{algorithm}%
\usepackage{algorithmicx}%
\usepackage{algpseudocode}%
\usepackage{listings}%
\usepackage{array}
\usepackage{acronym}
\usepackage[caption=false]{subfig}
\usepackage{anyfontsize}
\usepackage{xcolor}
\hypersetup{
    colorlinks=false
    }
\makeatletter
\AtBeginDocument{%
  \renewcommand*{\AC@hyperlink}[2]{%
    \begingroup
      \hypersetup{hidelinks}%
      \hyperlink{#1}{#2}%
    \endgroup
  }%
}
\makeatother

\begin{document}

\title[Article Title]{AI-assisted Advanced Propellant Development for Electric Propulsion}


\author[1]{\fnm{Angel} \sur{Pan Du}}

\author*[2]{\fnm{Miguel} \sur{Arana-Catania}}\email{humd0244@ox.ac.uk}

\author*[1,3]{\fnm{Enric} \sur{Grustan Gutiérrez}}\email{E.Grustan@cranfield.ac.uk}

\affil[1]{\orgname{Cranfield University}, \orgaddress{\street{College Road}, \city{Cranfield}, \postcode{MK43 0AL}, \country{United Kingdom}}}

\affil[2]{\orgname{University of Oxford}, \orgaddress{\street{Broad Street}, \city{Oxford}, \postcode{OX1 3BG}, \country{UK}}}

\affil[3]{\orgname{Ignition Space}, \orgaddress{\street{C/ Còrsega 384 6-1}, \city{Barcelona}, \postcode{08037}, \country{Spain}}}


\abstract{Artificial Intelligence algorithms are introduced in this work as a tool to predict the performance of new chemical compounds as alternative propellants for electric propulsion, focusing on predicting their ionisation characteristics and fragmentation patterns. 

The chemical properties and structure of the compounds are encoded using a chemical fingerprint, and the training datasets are extracted from the NIST WebBook. 

The AI-predicted ionisation energy and minimum appearance energy have a mean relative error of 6.87\% and 7.99\%, respectively, and a predicted ion mass with a 23.89\% relative error. In the cases of full mass spectra due to electron ionisation, the predictions have a cosine similarity of 0.6395 and align with the top 10 most similar mass spectra in 78\% of instances within a 30 Da range.}




\maketitle

\section{Introduction}

Despite the great potential of electric propulsion, the reliance on xenon (Xe) as the primary propellant presents challenges due to its rarity and escalating costs \cite{welle1993xenon}. As a consequence, a series of alternatives have been sought. The most direct approach has been using other noble gases such as krypton or argon; however, they present worse ionisation characteristics \cite{Andreussi2017, soton42236,oleson2004electri, plasma5030025, eckhaus2024student}, and in the case of krypton also a rapid cost increase \cite{osti_2377416}. Other atomic alternatives have also been explored in the p-block of the periodic table, such as bismuth or iodine, and while they can have excellent performance, they have condensation issues and, in the case of iodine, potential losses due to its molecular nature \cite{Szabo2017,szabo2013iodine,tverdokhlebov2001iodine}.
In light of these limitations, there is a growing interest in exploring alternative molecular compounds that can effectively substitute xenon and other chemical elements as a propellant for \ac{EP} while remaining stable \acs{EP} applications \cite{ling2020brief}.
Adamantane (C$_{10}$H$_{16}$) and buckminsterfullerene (C$_{60}$) are promising candidates, but not without drawbacks. Adamantane, along with iodine, has potential compatibility issues, such as spacecraft contamination and toxicity \cite{soton42236,oleson2004electri} and (C$_{60}$) temperature stability issues; elevated temperatures result in the fragmentation of the molecule, while reduced temperatures lead to resublimation onto the engine internal surfaces \cite{anderson1996fullerene, SCHARLEMANN2002865}.

While the highlighted candidates present some significant issues, the list of potential molecular propellants is practically limitless. However, the virtual infinity of the candidate list and the complexity of their characteristics make a trial-and-error approach impractical, especially when facing a new component without information on its characteristics (eg, ionisation energy or molecular stability). For this reason, other methodologies are currently under exploration, such as approaches that blend empirical experimentation with computational analysis \cite{Hill2005, Heinonen2008, Wolf2010, Wang2014, Krettler2021, DeProft1997, Flad2008, Gasteiger1992, Yang2021, Xu2020, Bremer2022}. Moreover, quantum chemistry calculations serve as the predominant method for accurately computing various fragmentation parameters. These calculations typically require inputs related to the molecular structure and electronic configuration of the compounds under investigation \cite{Allen2014, Cautereels2016,   Jonkman1974}. Nonetheless, these strategies often necessitate significant temporal and computational investments, accompanied by the challenge of potential inaccuracies due to reliance on individual expertise.

To overcome similar or more complex challenges, \ac{ML} has become a mainstay of cheminformatics, especially for drug discovery \cite{LO20181538}. Therefore, the implementation of \acs{ML} techniques holds promise for efficiently and accurately identifying ideal \acs{EP} propellant candidates. By leveraging \acs{ML} techniques, this project aims to streamline the selection process, reduce reliance on expensive propellants, identify viable alternatives that offer comparable performance characteristics, and kickstart the development of mission-tailored propellants by providing a tool capable of predicting the behaviour of novel molecular compounds when used as EP propellants even when the only information available is the compound structure.

\section{Methodology}

\subsection{Relevant physical parameters}
When facing a potential brand-new propellant, a wide array of properties can be of interest, such as density, toxicity, corrosiveness, or physical state at ambient temperature. However, this study focuses on properties relevant to the ionisation of molecular compounds.

Probably some of the obvious parameters affecting the performance of \acs{EP} systems are the ionisation energy and molecular mass of the propellant \cite{Choueiri2009}. To minimise the energy required for generating a high-density plasma, an ideal propellant should possess a low ionisation energy and a high ionisation cross-section \cite{Mazouffre_2016}. The ionisation energy represents the minimum energy necessary to remove an electron from an atom or molecule, transforming it into a charged ion. Conversely, the ionisation cross-section quantifies the likelihood of an ionising collision occurring between a charged and neutral particle \cite{Choueiri2009}.
And while the current power levels available to \acs{EP} promote heavy, easily ionised propellants (such as Xe), which deliver higher momentum, at the cost of a smaller specific impulse, the optimal molecular mass of the propellant is a new spin on the classical problem of Isp optimisation \cite{Jahn2006}, depending on the mission  \(\Delta\)v and power system mass efficiency.

Unlike atomic propellants, the possibility of splitting a molecule makes other parameters relevant. One of them is the minimum appearance energy, which represents the minimum energy required for a molecule to ionise through fragmentation. A priori, the preferred attribute for appearance energy is high, as it implies greater molecular stability, reducing the likelihood of decomposition upon ionisation. However, in some cases, if the appearance energy is too low, the propellant molecules might disintegrate prematurely during ionisation, resulting in inefficient thrust production and potential harm to the propulsion system \cite{anderson1996fullerene}.

In addition to the stability assessment, it is critical to assess how the molecules will fragment. The ion prediction from minimum appearance energy is the expected mass of the ion resulting from the molecule fragmentation at that energy, thus giving an idea of how the molecule fragments and, consequently, the potential losses due to unused mass or low polydispersive efficiency (acceleration losses due to a diverse ion mass or specific charge population). So using Buckminsterfullerene, C$_{60}$, as an example, at 7.8 eV the whole molecule will ionise and at an appearance energy of 20.2 eV the ion C$_{58}^+$ will appear (as will C$_{2}$) due to the molecule breaking, the mass of the ion in this case will be 696.62 Da \cite{C60}. If the ionisation energy increases, smaller ions will appear. 

For these reasons, \ac{IE}, minimum \ac{AE}, and ion prediction from the minimum \acs{AE} fragmentation are the chosen parameters to test the ML algorithms for studying molecular \acs{EP} propellants. These three general parameters, plus the molecular mass, provide a general idea of the quality of a compound as a propellant. As a first approximation, the propellant should be easy to ionise (low \ac{IE}), stable (high \ac{AE}), and in the case of creating ions also via fragmentation, the ion mass should be close to the original one (higher polydispersive efficiency). 

While the previous parameters offer a general idea of the quality of a propellant for plasma thrusters, the ionisation and fragmentation phenomena are more complex, creating a distribution of charged ions. Similarly to the case of a high-power \ac{HET} with multiply ionised ions, a range of specific charges in the plasma beam can result in losses that need to be accounted for \cite{Goebel2023}, in particular if the specific charges are very dissimilar \cite{Lozano}. Mass spectrometry analysis provides valuable insight into the fragmentation patterns, which can help assess the stability and performance of potential propellants and can be used to evaluate losses due to the distribution of specific charges in the plasma beam (polydispersive efficiency) \cite{Grustan2017}. The ideal case is a monodisperse beam; failing this, lowering the coefficient of variation of the specific charge improves the polydispersive efficiency \cite{Grustan2017}. In this case, mass spectra offer the complete fragmentation profile for a given ionisation energy. Particularly, a focus is placed on \ac{EI} \ac{MS}, due to its widespread use in the field of \acs{EP}, providing valuable data for predicting the fragmentation patterns of chemical compounds \cite{soton42236, tverdokhlebov2001iodine, anderson1996fullerene}. 

Additionally, during the selection of alternative propellants for electric propulsion, other critical parameters should be taken into account, such as chemical compatibility with both the system \cite{soton422369,oleson2004electric} and the thruster acceleration mechanism \cite{Choueiri2009}, as well as scalability with power \cite{Wittenberg2019}. However, these parameters are system-specific and depend on the trade-offs between propellant performance and its system engineering impacts, and while ML can also be used to optimise a whole system and identify other characteristics such as toxicity, we chose to focus on the prediction of the total mass spectra to showcase the potential of ML to help predict more complex (and system-specific) problems.

\subsection{Molecular structure encoding and data}

As discussed, molecule fragmentation during ionisation is a significant part of the study. Consequently, inputs related to the molecular structure and electronic configuration of the compounds under investigation are needed. Although different methods can be used to provide this information, such as bond characteristics, electronic configuration, harmonic vibrational data, or mass spectrum \cite{Qu2013, Li2021, Ljoncheva2022, Matthews2020, Yerokhin2020, Zeng2019}, the molecular fingerprint is the more direct and flexible method to provide input to the \acs{ML} algorithms as it does not require of extra computational or experimental steps \cite{LO20181538, Klamt2012, Krettler2021, Qiu2023, Stewart2023, Zhang2022}. The molecular fingerprint is a multidimensional vector containing the molecule's atomic elements and structure.

For this simplicity and directness, fingerprints are chosen as the primary input of the models, in particular, the \ac{ECFPs}, a form of molecular representation widely used in computational chemistry. They are generated through an algorithm known as the Morgan algorithm \cite{Morgan1965}, which is rooted in graph theory and captures the structural features of chemical compounds. \acs{ECFPs} are binary strings that encode the presence or absence of specific substructures within a molecule based on the neighbouring atoms and bonds within a defined circular radius.

The generation of these fingerprints was accomplished using the \textit{RDKit Cheminformatics} package \footnote{\url{https://www.rdkit.org/}}, with a chosen fingerprint length of 4096 and a radius of 2. These specific values were chosen based on considerations of both efficiency and representational capacity.

A longer fingerprint length enhances the depiction of molecular structure by accommodating a wider array of distinctive substructures at the cost of computational requirements and memory consumption. A larger radius captures more atoms and bonds, enriching the portrayal of molecular interactions and correlations. Nonetheless, an excessive radius risks including extraneous information, potentially introducing noise to the fingerprint representation.

Furthermore, an exploration of additional inputs, as well as some adjustments to these, was undertaken to potentially enhance the models' performance. As an example, the inclusion of the mass data from the chemical compounds and the adjustment of fingerprint lengths and radii were among the investigated modifications. The efficacy of these additional inputs will be elaborated upon and scrutinised in the subsequent results section.

As is usual in \acs{ML}, the choice of the dataset to train the algorithms is critical. Thanks to the ready availability of some of the aforementioned data in the \ac{NIST} Chemistry WebBook \footnote{\url{https://webbook.nist.gov/chemistry/}}, along with its widespread use in literature, makes it an ideal choice for this study. Chemical data can often be an expensive endeavour. However, the \acs{NIST} Chemistry WebBook offers a reduced version accessible to everyone, encompassing up to 72,618 compounds at present.

\begin{figure}[htb]
    \centering
    \subfloat[\acs{EI} \acs{MS} database\label{fig:mass_distribution_MS}]{
        \includegraphics[width=40mm]{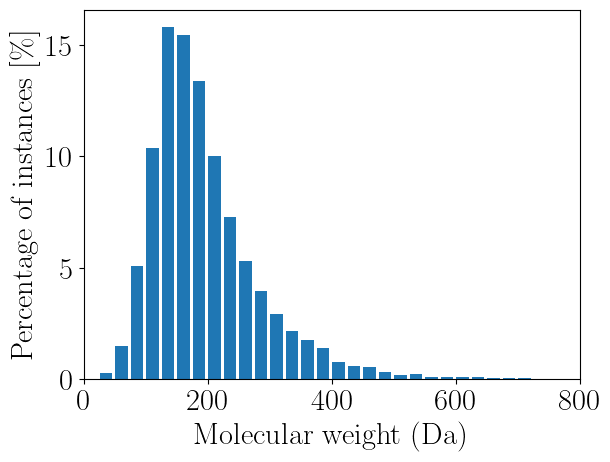}
    }
    \hfill
    \subfloat[\acs{IE} database\label{fig:mass_distribution_IE}]{
        \includegraphics[width=40mm]{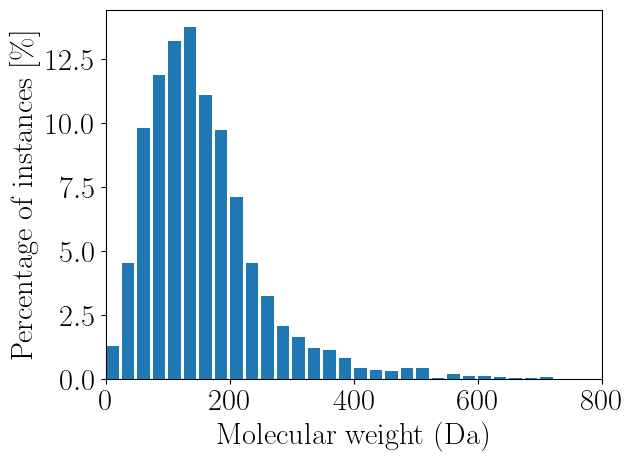}
    }
    \hfill
    \subfloat[\acs{AE} database\label{fig:mass_distribution_AE}]{
        \includegraphics[width=40mm]{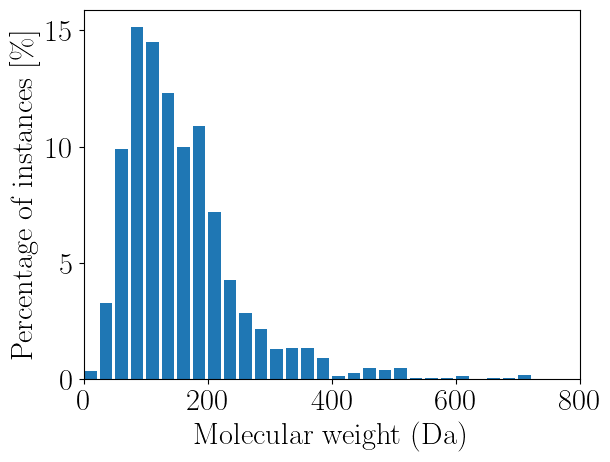}
    }
    \caption[Distribution of molecular weight across databases]{Distribution of molecular weights of chemical compounds in the three databases. The histograms indicate the percentage of compounds in specified ranges of molecular weight.}
    \label{fig:mass_distribution}
\end{figure}

\begin{figure}[htb]
    \centering
    \subfloat[\acs{IE} prediction\label{fig:IE_distr}]{
        \includegraphics[height=27mm]{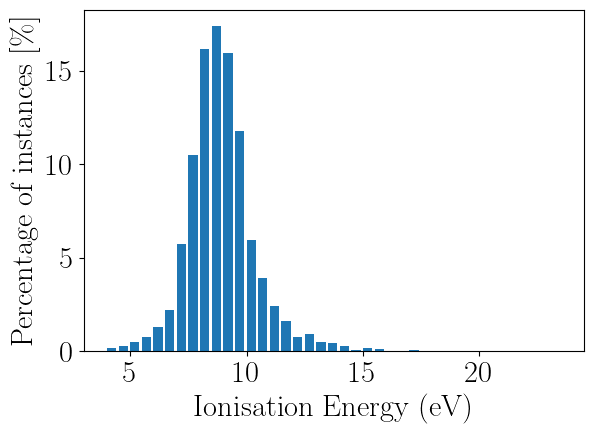}
    }
    \hfill
    \subfloat[\acs{AE} prediction\label{fig:AE_distr}]{
        \includegraphics[height=27mm]{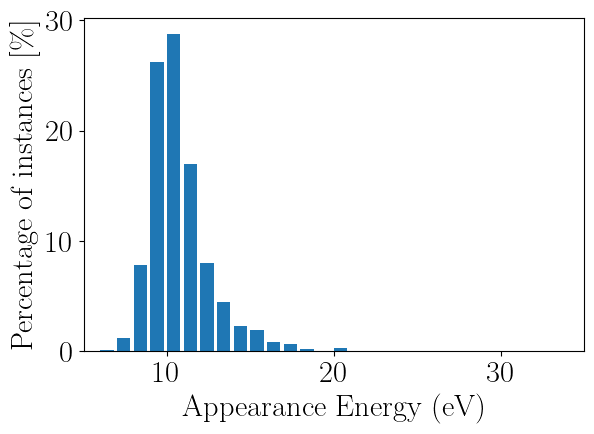}
    }
    \hfill
    \subfloat[\acs{AE} ion prediction\label{fig:AEion_distr}]{
        \includegraphics[height=27mm]{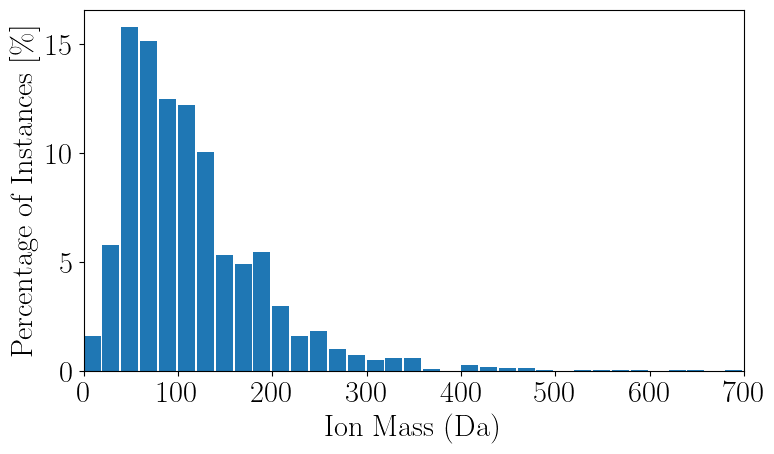}
    }
    \caption{Distribution of the prediction parameters in the datasets.}
    \label{fig:energies_distribution}
\end{figure}

Ultimately, data scrubbing from the \acs{NIST} database yielded mass spectra data from 21,142 distinct compounds, encompassing all available content regarding this parameter. In addition, ionisation energy data from 3,073 compounds and minimum appearance energy data from 2,148 compounds were successfully obtained. The primary drawback of open-access data collected in this database is its emphasis on relatively low-mass chemical compounds (50-200 Da). Further details on the molecular weight distribution for the three databases are available in Figure \ref{fig:mass_distribution}. Moreover, the ranges analysed for the \acs{IE} and \acs{AE} are 3.89-24.59 eV and 5.06-36.00 eV, respectively, and for the ion mass derived from the \acs{AE} is 1-698 Da. For better visualisation, the distributions of these parameters in their datasets are represented in Figure \ref{fig:energies_distribution}

\subsection{Model architectures and training details}
\label{m_train}

For this project, the chosen \acs{ML} model to produce the predictions is based on neural networks, specifically utilising a \ac{MLP} structure for all prediction scenarios, and additionally using \ac{LSTM} and bidirectional long short-term memory (Bi-\acs{LSTM}) networks for the case of mass spectrum prediction. The decision to use an \acs{MLP} is grounded in its successful implementation in reference papers for mass spectrum prediction \cite{Zhang2022, Wei2019}, ensuring comparability with prior research and establishing a reliable and consistent approach in evaluating the current \acs{ML}-based assessment system. 

Nevertheless, there are critical differences between the present work and these references. Although previous research has focused on the use of machine learning models to predict mass spectra, this article extends the study further by incorporating additional predictive tasks, such as \acs{IE} and minimum \acs{AE} energies, thus expanding the scope of the analysis. Moreover, unlike the previously discussed research that utilised a sub-database available only for the licensed version of \acs{NIST}, the current study relies exclusively on the open-access version. Therefore, straightforward comparisons cannot be made in terms of identical evaluation metrics. However, a similar methodological configuration exists with regard to measuring model performance, as discussed in subsequent sections.

Regarding the aforementioned \acs{ML} models, firstly, a multilayer perceptron  \cite{bengio2017deep} is a neural network composed of layers of neurons where each neuron of a layer is connected to all neurons of the following layer. The neurons represent a linear combination of the outputs provided by the connections of the previous layer followed by a non-linear function.
 
A long short-term memory \cite{hochreiter1997long} is a type of recurrent neural network. Unlike the MLP, which processes each input data point independently to produce an individual output, the latter keeps an internal memory of the processing of each input which is used for processing the following one. This allows it to relate them and thus deal with sequential data. For example, when processing a sequence of atoms, for each atom, it uses the information of all previous atoms in the sequence. The LSTM has a complex internal structure, combining multiple types of linear and non-linear functions in internal operations called gates, which allow the network to learn how to manage the internal memory for each specific task.
 
A Bi-\acs{LSTM} \cite{graves2005framewise} is a recurrent neural network which combines two LSTM networks. Each of them processes the input sequence in a different direction, starting from a different end of the sequence. This design is appropriate for sequences where each element is related to both previous and posterior elements of the sequence.

The division of the dataset was performed using a 90-5-5 randomised split, where 90\% was allocated to the training set, and 5\% to the validation set and to the test set, as suggested in \cite{Zhang2022, Wei2019}. The training set is utilised to fit the model, the validation set is employed to fine-tune hyperparameters and prevent overfitting, and the test set is reserved for evaluating the final model's performance on unseen data. For this reason, performance metrics such as loss and relative error are meant to differ between the test subset, and training and validation. Furthermore, each model training is different, meaning that the loss landscape is defined specifically by the data provided \cite{Li2018}. Hence, divergent behaviors can be found within the training losses from distinct predictions.

The loss function employed was SmoothL1Loss. Tuning of hyperparameters was conducted for the present studies considering various activation functions, such as leaky \ac{ReLU}, tanh, and sigmoid, different optimisers (e.g., \ac{SGD}, Adagrad, Adadelta, and RMSprop) \cite{he2016survey}, as well as other relevant parameters for the training. The alternative proposed activation functions and optimisers were selected based on their common usage and effectiveness in machine learning literature \cite{Goodfellow2016, LeCun2015}, and are further discussed in the corresponding subsections from their respective prediction.

\subsection{Ionisation energy prediction}
\label{IE_pred}

Like in all the following predictions, molecular fingerprints were employed as input data for the model. This is due to the accuracy of the molecular fingerprint in representing the compound's structure, a factor commonly used in quantum chemical calculations for ionisation energy determination. Additionally, insights from \cite{Stewart2023}, focusing on ionisation energy prediction for volatile organic compounds (VOCs) using molecular fingerprints and neural network models, further support the effectiveness of this methodology.
 
Thus, the architecture from the model consists of an \acs{MLP} structure, with appropriate hyperparameters properly tuned to optimise the results, whose outputs are the minimum appearance energy achieved through the electron ionisation method.

The relative error was chosen to evaluate the effectiveness of the \acs{MLP} algorithm, providing a straightforward and interpretable measure of the prediction's accuracy. This value represents the mean relative error, calculated from each compound, considering the predicted parameter with respect to the real one.

\subsection{Minimum appearance energy prediction}
\label{AE_pred}

The same approach was adopted for the minimum appearance energy prediction, carefully tuning hyperparameters to optimise the results. 

Unlike ionisation energy, no previous works were found regarding potential inputs or the training model. However,  given the success found for the approach detailed in the preceding section, a molecular fingerprint input and a \acs{MLP} architecture were chosen. Hence, the desired outcome of this architecture aims to determine the minimum appearance energy through the \acs{EI} method.

Once again, the relative error was selected as the evaluation metric since the minimum appearance energy prediction, like the ionisation energy prediction, generates a single output.

\subsection{Ion mass prediction from the minimum \acs{AE} fragmentation}
\label{AEion_pred}

For the prediction of the ion generated due to fragmentation at the minimum appearance energy, the lack of preceding references necessitated an innovative approach. In this regard, the molecular fingerprint was chosen as the primary input for the training model, and the \acs{MLP} as the architecture, leveraging their successful track record in the previous predictions. The fitness metric used is the relative error of the molecular mass of the predicted ion.

\subsection{Mass spectrum prediction}
\label{MSpred}

The mass spectra prediction for chemical compounds was also based on \acs{MLP} using a similar procedure to \cite{Zhang2022} and \cite{Wei2019}. However, additional architectures were also explored to enhance the results. Among these, \acs{LSTM} network and bi-\acs{LSTM} network were considered due to their potential advantages in handling sequential data \cite{Le2019}.

These studies have intimated the existence of correlations between consecutive peaks in the mass spectrum. This characteristic adjusts well with the inherent capability of this architecture to capture sequential dependencies in data. By utilising \acs{LSTM}, the algorithm aims to exploit the information encoded in the sequential order of peaks, enhancing its ability to discern patterns and relationships within the mass spectrum data. This aligns with the fundamental motivation behind adopting \acs{LSTM} for this specific prediction task.

In this case, the input layer receives the molecular fingerprint data, more specifically, the \acs{ECFPs}, and the output layer generates the predicted mass spectra. This final layer of the model possesses an equal length to the desired output, in other words, the mass spectra vector length. This vector was structured with relative intensities, which were positioned in the vector according to their respective mass-to-charge ratio. Moreover, an attempt was made to enhance the model's performance by normalising the values with respect to the peak situated at 100\%.

Following the reference papers, the cosine similarity metric was utilised to assess the performance of their model (calculating the cosine of the angle between the predicted mass spectra vector and the one in the database).

Since the objective is to design a model capable of accurately predicting the electron ionisation mass spectrum for any given molecule, the model is then employed to construct an augmented reference library comprising both predicted and experimentally measured spectra. 

\begin{figure}[htb!]
    \centering
    \includegraphics[width=120mm]{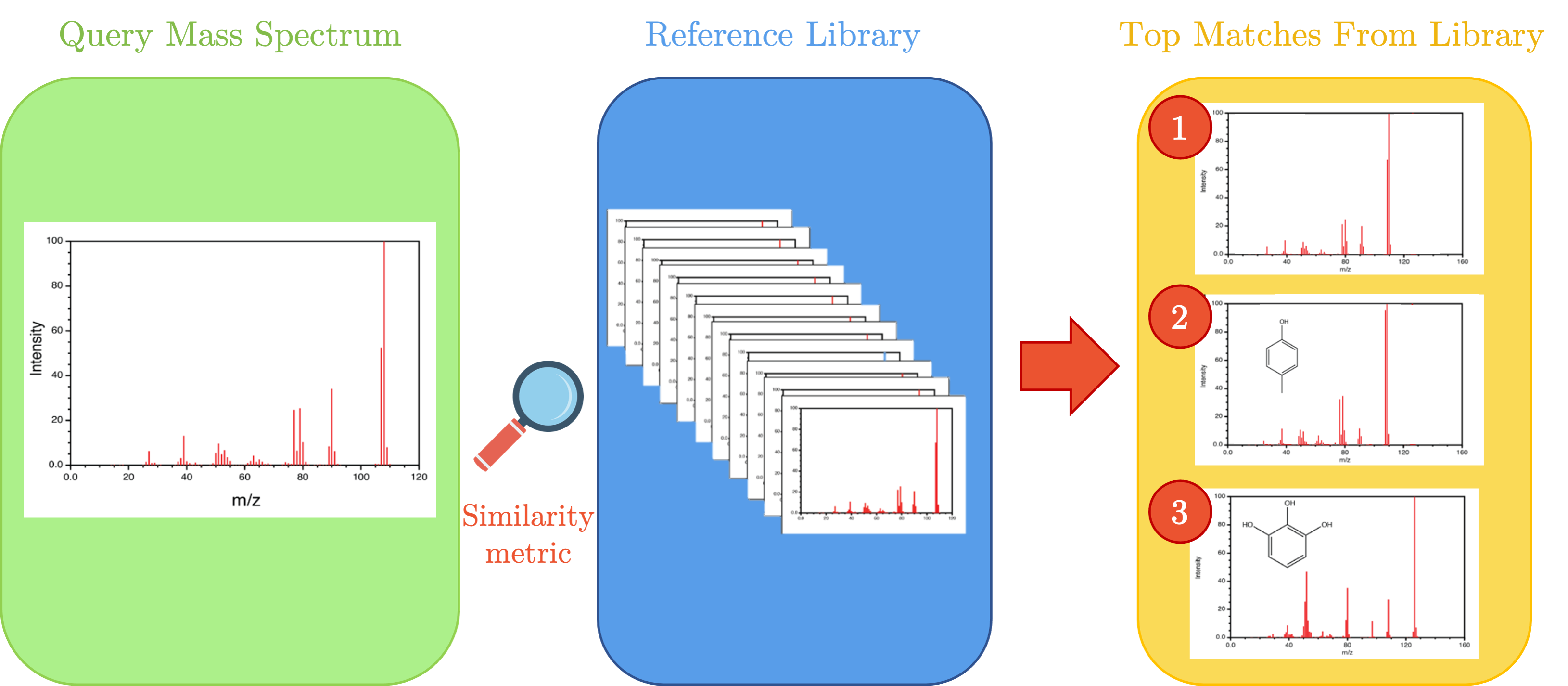}
    \caption{Overview of the library matching. Modified from \cite{Wei2019}.}
    \label{fig:accuracy}
\end{figure}

Subsequently, library matching is performed: the cosine similarity between each predicted spectrum and every spectrum from the augmented library is calculated and sorted from highest to lowest. Then, the rank of the correct spectrum is recorded (for each molecule used as an input during validation). The results obtained in this section are also evaluated using recall@k, where k represents values of 1, 5, or 10. This parameter measures the proportion of cases in which the correct spectrum appears within the top k ranked results. For instance, recall@1 indicates the percentage of times the correct match is ranked first, while recall@10 reflects how often it appears within the top ten.  Figure \ref{fig:accuracy} shows a visual depiction of this process.

Furthermore, a mass filter, similar to the one employed in the previous papers, was also integrated into the algorithm to facilitate the similarity search and analyse a more concise accuracy. The mass filter excludes species that fall outside the actual compound mass within a predetermined tolerance of 30 Da ($\pm$15 Da). This mechanism ensures that only compounds within the specified mass range are taken into consideration. 

This strategic filtering was designed to retain a substantial number of potential candidates while minimising the reduction in the candidate list. The chosen tolerance value was determined through careful analysis, ensuring an average subset size of approximately 400 compounds within the filtered dataset. This value was selected to maintain a relative proportion to the filtered dataset length observed in the reference papers \cite{Zhang2022, Wei2019}.

\section{Results and discussion}

In this section, the results and analysis obtained from the application of the \acs{ML}-based assessment system for mass spectrum prediction, ionisation energy prediction, and minimum appearance energy prediction are presented. 

The summary of the optimal configuration hyperparameters found during the hyperparameter tuning of the various machine learning models is presented in Table \ref{tab:hyper}.

\begin{table}[htb!]
\centering
\caption{Configuration parameters summary table for optimal prediction results} 
\label{tab:hyper}
\begin{tabular*}{\textwidth}{@{\extracolsep\fill}ccccc}
\hline
                                                                               & \textbf{\acs{IE}}    & \textbf{\acs{AE}}    & \textbf{\acs{AE} mass} & \textbf{MS}\\ \hline
\textbf{MLP layers}          & 5            & 5              & 5              & 5                \\
\textbf{Hidden neurons}                                                              & 512            & 64             & 64  & 4096              \\
\textbf{Epochs}                        & 22             & 28             & 16     & 28           \\
\textbf{Batch size}                                                                    & 2              & 2              & 1    & 32             \\
\textbf{Learning rate}               & 0.001          & 0.01           & 0.04       & 0.001      \\
 \textbf{Dropout}                                                                      & 0.15           & 0.3            & 0.4      & 0.2         \\
\textbf{Activation function}    & ReLU           & Sigmoid        & Leaky ReLU     & Leaky ReLU  \\
 \textbf{Optimiser}                                                                   & RMSProp        & Adagrad        & Adagrad       & Adam    \\   \hline
\end{tabular*}
\end{table}

\subsection{Ionisation energy prediction results}

\begin{figure}[htb]
    \centering
    \begin{minipage}{.5\textwidth}
        \centering
        \includegraphics[width=60mm]{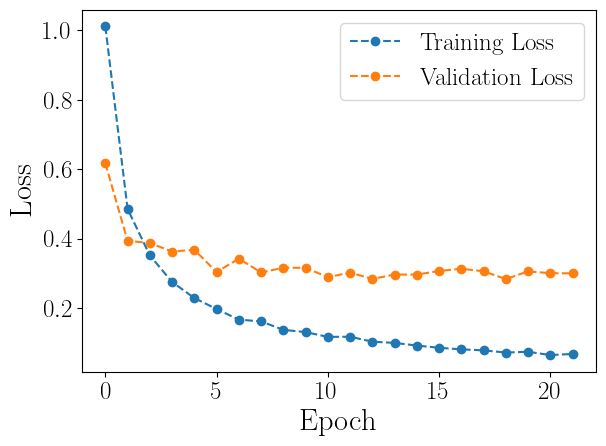}
        \label{fig:lossIE}
    \end{minipage}%
    \begin{minipage}{0.5\textwidth}
        \centering
        \includegraphics[width=60mm]{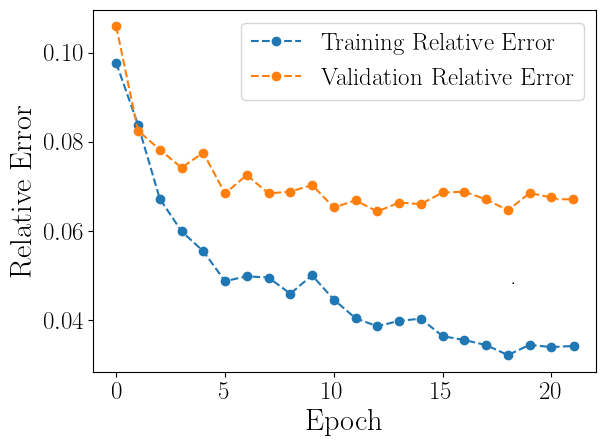}
        \label{fig:reIE}
    \end{minipage}
    \caption[Performance from the IE prediction model]{Performance from the IE prediction model with the configurations from Table \ref{tab:hyper} using the training and validation sets.}
    \label{fig:IE_curves}
\end{figure}

The best predictions of the ionisation energy using the methodology of Section \ref{IE_pred} are obtained using the configuration succinctly summarised in Table \ref{tab:hyper}.
Figure \ref{fig:IE_curves} offers the evolution of the training and validation loss after each epoch (a measurement of the training effectiveness after each training round), illustrating a consistent downward trend across all curves, successfully aligning with the expected behaviour. 

The curve discrepancy and the validation loss curve noise relative to the smoother training curve could likely stem from the relatively modest database size and the potential for further training gains. Nonetheless, the overarching trajectories converge to a stable value for both training and validation losses and for the relative error. In fact, for this optimised model, a successful prediction of ionisation energy was accomplished, with a considerably low relative error of 6.87\%. In comparison, if the average ionisation energy of the whole population was used as a predictor, the average relative error would be 18.05\%. 

It is important to note that the ionisation energies in the test set are representative of the whole set, falling within 69.48\% of the average value, with values ranging from 3.9 to 21.6 eV. The absolute error of the predictions was analysed as a function of the energy, and no trend was found between the two.

In Table \ref{tab:resultsIE}, the 5 predictions with the lowest relative error are presented.

\begin{table}[htb]
\centering
\caption{Top 5 predictions from the \acs{IE} prediction with the model from Table \ref{tab:hyper}}
\begin{tabular}{cccccc}
\hline
 & { $\mathrm{C_8H_6F_2O_2}$} & {$\mathrm{C_{13}H_{11}N}$} & {$\mathrm{C_{15}H_{12}}$} & {$\mathrm{C_{10}H_{13}BrO}$} & {$\mathrm{C_{12}H_{16}O}$} \\ \hline
\textbf{Predicted \acs{IE} [eV]} & 8.88 & 8.15 & 7.70 & 8.55 & 7.99 \\ 
\textbf{Real \acs{IE} [eV]} & 8.88 & 8.15 & 7.70 & 8.54 & 8.00        \\ \hline
\end{tabular}
\label{tab:resultsIE}
\end{table}

\subsection[Minimum \acs{AE} prediction results]{Minimum appearance energy prediction results}

This section delves into the outcomes of forecasting the minimum appearance energy. Firstly, attention is directed towards Table \ref{tab:hyper}, which exhibits the configurations of the models that produced the least relative errors reached, pertaining to the prediction of minimum appearance energy.

\begin{figure}[htb]
    \centering
    \begin{minipage}{.5\textwidth}
        \centering
        \includegraphics[width=60mm]{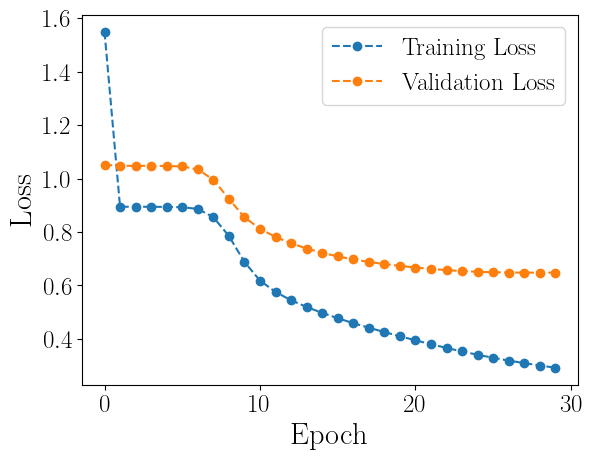}
        \label{fig:lossAE}
    \end{minipage}%
    \begin{minipage}{0.5\textwidth}
        \centering
        \includegraphics[width=60mm]{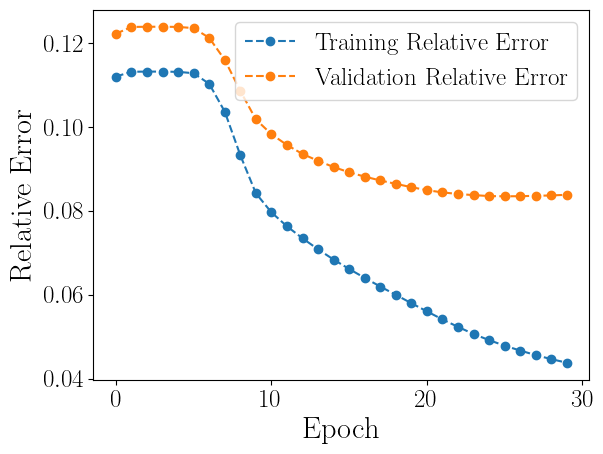}
        \label{fig:reAE}
    \end{minipage}
    \caption[Performance from the minimum \acs{AE} prediction model]{Performance from the minimum \acs{AE} prediction model with the configurations from Table \ref{tab:hyper}.}
    \label{fig:AE_curves}
\end{figure}

Table \ref{tab:hyper} exhibits the configuration of the model that produced the lowest relative errors for the prediction of minimum appearance energy.
Employing these identified hyperparameters and settings for the model, the loss curves and relative errors of Figure \ref{fig:AE_curves} were obtained.

The graphs consistently portray a common trend, with all curves gradually descending and converging towards stable values. 
As in the previous section, the validation curves stabilise to a higher value than the training ones but with a very low relative error, but the convergence criteria must also consider the validation subset results to avoid overfitting. Moreover, there is an initial plateau in these curves, which can be potentially attributed to a less convex loss landscape with respect to the preceding section.

Ultimately, employing these identified hyperparameters and settings for the model, the attained minimal relative error for the prediction of minimum appearance energy yielded a value of 7.99\%. By calculating the relative error using the mean value from the dataset, the error increases to 19.1\%, thereby reinforcing the accuracy of the model. All appearance energies values from the test dataset range within 44.09\% of the average value, in the 7.1-18.3 eV range. In this case, the error also did not have any trend with respect to the energy. The 5 best predictions are represented in Table \ref{tab:resultsAE}.

\begin{table}[htb]
\centering
\caption{Top 5 predictions from the \acs{AE} prediction with the model from Table \ref{tab:hyper}}
\begin{tabular}{cccccc}
\hline
 & { $\mathrm{C_3H_5ClO}$} & {$\mathrm{C_{4}H_{8}OS}$} & {$\mathrm{C_{7}H_{7}F}$} & {$\mathrm{C_{7}H_{16}O_2}$} & {$\mathrm{C_{2}H_{7}N}$} \\ \hline
\textbf{Predicted \acs{AE} [eV]} & 10.29 & 9.89 & 11.88 & 10.32 & 9.53 \\ 
\textbf{Real \acs{AE} [eV]} & 10.29 & 9.90 & 11.90 & 10.30 & 9.55        \\ \hline
\end{tabular}
\label{tab:resultsAE}
\end{table}

\subsection[Ion mass prediction results]{Ion mass prediction from the minimum \acs{AE} fragmentation results}
\label{AEionres}

\begin{figure}[htb]
    \centering
    \begin{minipage}{.5\textwidth}
        \centering
        \includegraphics[width=58mm]{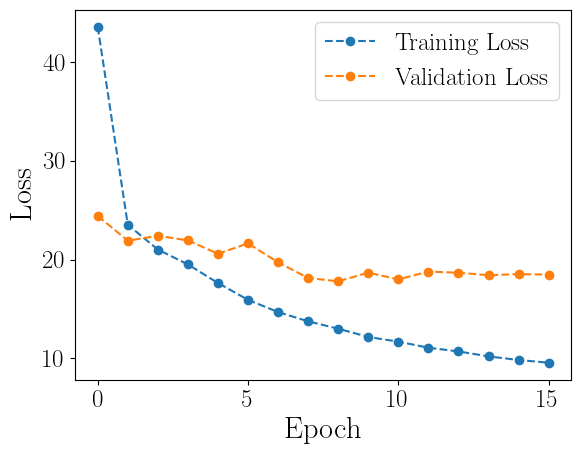}
        \label{fig:lossAEion}
    \end{minipage}%
    \begin{minipage}{0.5\textwidth}
        \centering
        \includegraphics[width=60mm]{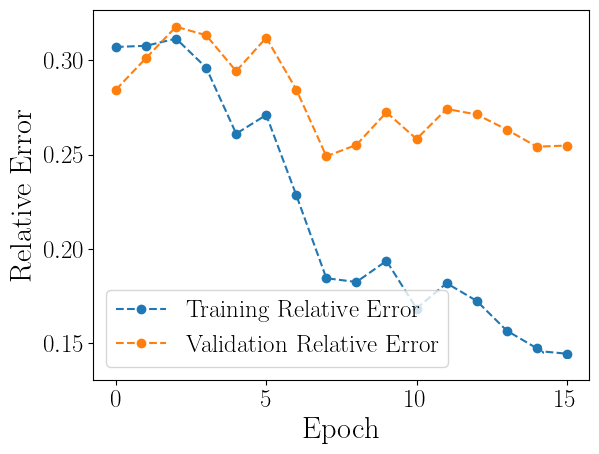}
        \label{fig:reAEion}
    \end{minipage}
    \caption{Performance from the prediction model of the ion mass for minimum \acs{AE} with the configurations from Table \ref{tab:hyper}.}
    \label{fig:AEion_curves}
\end{figure}

The model configurations that yielded the lowest relative errors are depicted in Table \ref{tab:hyper}. Using those parameters, the training and ion mass prediction relative error curves of Figure \ref{fig:AEion_curves} were obtained.

In contrast to all the previous results, the ion mass prediction graphs display a comparatively higher degree of noise, ultimately converging to relatively higher error rates. This is not necessarily a surprising fact, as the algorithm needs to predict not only when the molecule will break (implicitly, the appearance energy is also necessary) but also how it will break. Therefore, the molecule's chemical structure (and its encoding) is much more critical, increasing the problem's complexity.

Since these results exhibited a less satisfactory performance, some attempts were made to enhance the model. Firstly, the normalisation of the outputs was tested, involving the prediction of the ion mass percentage relative to the initial mass. Furthermore, other models, such as LSTM, were also implemented. However, this adjustment did not yield improvements in model outcomes. 

Additionally, the incorporation of the original compound mass as an input led to some enhancements in the model. Consequently, normalising this mass input with respect to the maximum value across the database was explored, given that the remaining inputs, which are the \acs{ECFPs} are represented as binary bits (0 or 1). Despite these measures, improvements remained elusive. Finally, the relative error for the associated ion mass prediction culminated at 23.89\%, with the data subset presenting a wide range of values from 6-625 Da, that is a 98.10\% variation from the average value. Taking the average value from the entire dataset, a relative error of 96.74\% would be obtained, so while the relative error is higher than for the previous parameters, the algorithm still outperforms the average significantly.
In Table \ref{tab:resultsAEmass}, the best 5 predictions from this model are listed.

\begin{table}[htb]
\centering
\caption{Top 5 predictions from the ion mass for minimum \acs{AE} with the model from Table \ref{tab:hyper}}
\begin{tabular}{cccccc}
\hline
 & { $\mathrm{C_9H_{12}}$} & {$\mathrm{C_{13}H_{21}NO}$} & {$\mathrm{C_{7}H_{8}}$} & {$\mathrm{C_{9}H_{1}ClN_2}$} & {$\mathrm{C_{7}H_{5}NO_4}$} \\ \hline
\textbf{Predicted ion mass [Da]} & 57.12 & 105.16 & 45.06 & 91.13 & 78.11 \\ 
\textbf{Real ion mass [Da]} & 57.67	& 104,05	& 46.01	& 94.45 &	82.06
        \\ \hline
\end{tabular}
\label{tab:resultsAEmass}
\end{table} 

Further iterations involving variations in the fingerprint inputs, such as alterations to radii and bit string length, were also trialled but yielded no enhancements either. Thus, it is plausible that the need for additional inputs or the inherently variable nature of the parameter under prediction could cause this to be a complicated value to predict.

\subsection{Mass spectrum prediction results}

Finally, three different model architectures were used to predict the full mass spectra resulting from the electron ionisation of a molecule with an energy of 70 eV, the most complex problem. This specific value from the excitation energy is standardised for this mass spectrum technique. Through hyperparameter tuning, it was determined that the \acs{MLP} proved to be the most effective model for predicting the mass spectrum, achieving a notable cosine similarity of 0.6395 and a recall@10 of 60.68\%, with its optimal hyperparameters.

In contrast, the \acs{LSTM} model yielded a cosine similarity of 0.5517 and a recall@10 of 49.62\% with the best-tuned parameters, while the bi-\acs{LSTM} demonstrated a cosine similarity of 0.5708 and a recall@10 of 52.36\%.

The inclusion of the mass from the chemical compounds as an input and the variation of the input features, such as altering \acs{ECFPs}, in terms of bit length or radii, did not yield performance improvements. Nevertheless, a notable enhancement was observed through the normalisation of relative intensities from the mass spectrum data.

Table \ref{tab:hyper}, details the specific values for the different hyperparameters that yielded the optimal results. 

\begin{figure}[htb]
    \centering
    \begin{minipage}{.5\textwidth}
        \centering
        \includegraphics[width=60mm]{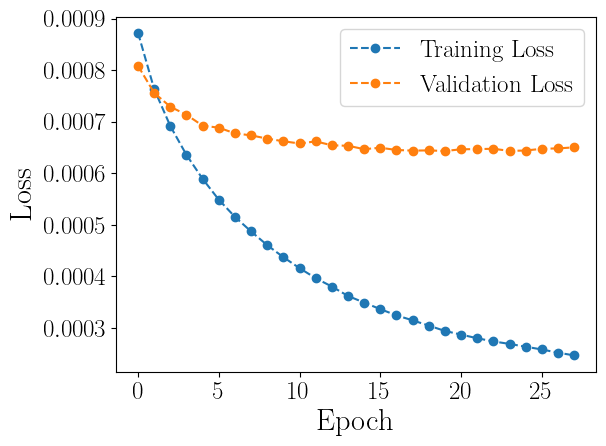}
        \label{fig:ms_loss}
    \end{minipage}%
    \begin{minipage}{0.5\textwidth}
        \centering
        \includegraphics[width=59mm]{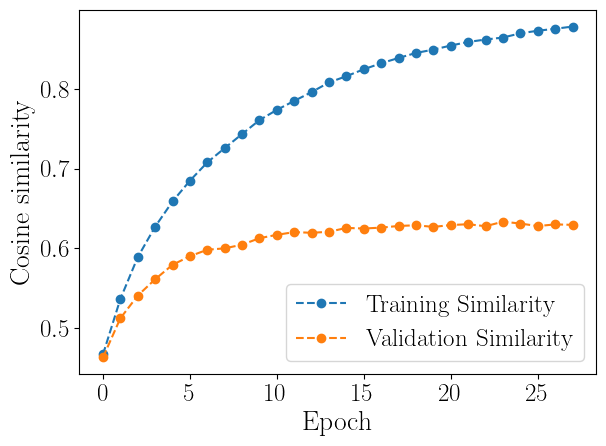}
        \label{fig:ms_sim}
    \end{minipage}
    \caption[Performance from the \acs{MS} prediction with MLP model]{Performance from the \acs{MS} prediction model with the configurations from Table \ref{tab:hyper}.}
    \label{fig:ms_curves}
\end{figure}

Visual representations of the loss function's behaviour over the training epochs and the corresponding trend in cosine similarity (Figure \ref{fig:ms_curves}) provide insights into the convergence behaviour and predictive accuracy achieved by the model from the previous table. 

Consequently, the final results, corresponding to the outcomes of the test subset, are displayed in Table \ref{tab:resultsMS}:

\vspace{1mm}
\begin{table}[htb]
\centering
\caption{Results from the \acs{MS} prediction with the model from Table \ref{tab:hyper}}
\begin{tabular}{cccc}
\hline
{ \textbf{Cos. sim.}} & { \textbf{recall@1}} & { \textbf{recall@5}} & { \textbf{recall@10}} \\ \hline
0.6395                                    & 31.04\%                                        & 52.89\%                                                                & 60.63\%                                                                 \\ \hline
\end{tabular}
\label{tab:resultsMS}
\end{table}
\vspace{-2mm}
Furthermore, with a mass filter to selectively focus on compounds with comparable masses, employing a tolerance of 30 Da, up to a recall@10 of 78\% was achieved. 

\begin{figure}[htb!]
    \centering
    \includegraphics[width=80mm]{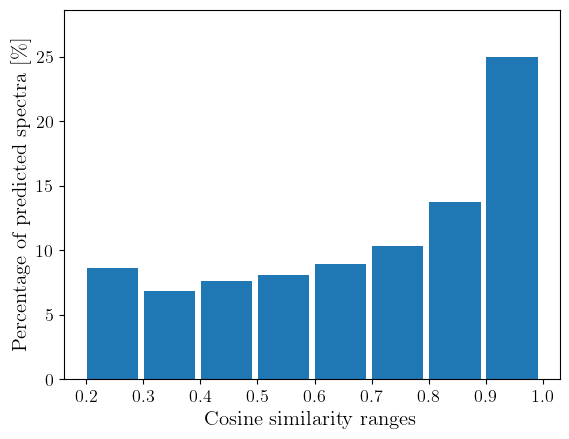}
    \vspace{-1mm}
    \caption[Cosine similarity distribution from the test set of the \acs{MS} prediction]{Cosine similarity distribution from the test set of the \acs{MS} prediction.}
    \label{fig:sim}
\end{figure}

To further gauge the efficacy of this model using the employed metrics, an additional illustrative representation is provided in Figure \ref{fig:sim}. This graph presents the distribution of predictions from the test subset compounds, with a mean cosine similarity of 0.6395 and a standard deviation of 0.2809, and are categorised according to specific ranges of cosine similarities. Each bar in the graph represents the percentage of compounds that fall within a particular range of the similarity score.

As perceptible from the graph, a substantial frequency of predictions demonstrates cosine similarity values surpassing 0.6, amounting to more than 60\% of occurrences. Furthermore, accurate predictions are discernible, as approximately one-fourth of the forecasts possess cosine similarity values exceeding 0.9.

\begin{figure}[htb]
    \centering
    \includegraphics[width=100mm]{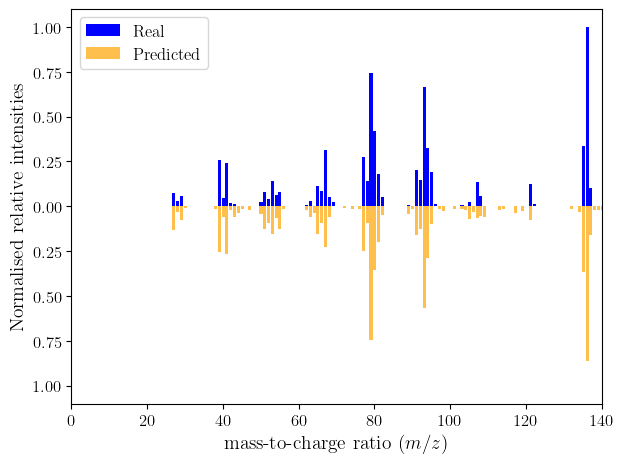}
    \vspace{-1mm}
    \caption[Comparison between the \acs{MS} prediction and the ground truth for C$_{10}$H$_{16}$]{Comparison between the \acs{MS} prediction and the ground truth for C$_{10}$H$_{16}$.}
    \label{fig:MS_ex2}
\end{figure}

\begin{figure}[htb!]
    \centering
    \includegraphics[width=100mm]{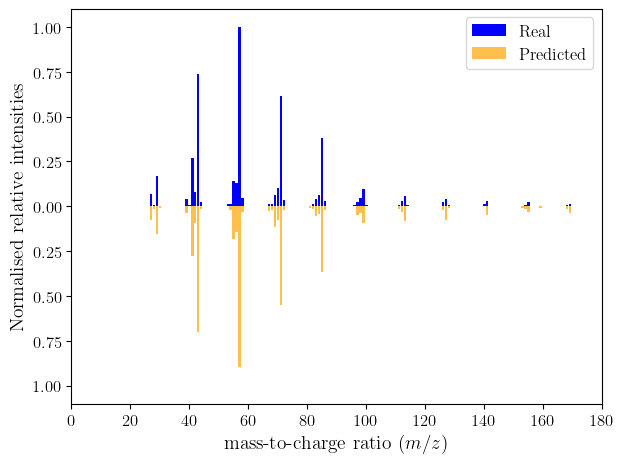}
    \vspace{-1mm}
    \caption[Comparison between the \acs{MS} prediction and the real for C$_{15}$H$_{32}$]{Comparison between the \acs{MS} prediction and the real for C$_{15}$H$_{32}$. Representation of the normalised relative intensities against the mass-to-charge ratio}
    \label{fig:MS_ex}
\end{figure}

Adamantane (C$_{10}$H$_{16}$), one of these cases and technologically relevant, is used to illustrate in Figure \ref{fig:MS_ex2} the ultimate outcomes of the prediction algorithm, and an additional example with pentadecane (C$_{15}$H$_{32}$) is also shown in Figure \ref{fig:MS_ex}. These compounds from their test subsets obtained cosine similarities of 0.9604 and 0.9912, respectively.

The graphical representation reveals a notable resemblance between the predicted outcomes and the actual values. While some disparities are discernible, particularly in the case of lower peaks, the accuracy of prediction is remarkable, especially in regard to the more prominent peaks. This alignment is particularly crucial as these significant peaks typically encapsulate substantial information within the mass spectrum.

Similarly to the case of ion mass prediction, the interaction between chemical bonds within the molecule reduces the fingerprint's effectiveness as a unique algorithm input. While the tuning of the fingerprint has not yielded any significant improvement, changing the algorithm (and fingerprints) depending on the number of bonds might be an improvement venue. 

Furthermore, the predictive performance has the potential for enhancement through the incorporation of a graph convolutional network (GCN) for processing the \acs{ECFPs}. GCN, as proposed in \cite{Wei2019} and successfully demonstrated in \cite{Zhang2022}, treats molecules as graphs, with atoms as nodes and chemical bonds as edges. This approach allows for a comprehensive representation of the entire molecular structure, potentially leading to improved predictive accuracy at a slightly higher computational cost. Nonetheless, it was not possible to recreate the same procedure from the references mentioned, due to the lack of specific sub-databases for it in the open-access version from \acs{NIST}.

Finally, one of the main challenges of this section is the training database. The \acs{MS} NIST database, as the name states, is used to calibrate mass spectrometers, and in consequence, its structure is not inherently optimized for training algorithms. Despite this, it remains the most suitable and comprehensive database currently available for our application. Three of the main limitations are that the spectra count the number of ions with a particular specific charge without reference to the number of ionised molecules, a similar consideration is that there is no data on ionisation efficiency (especially relevant for \acs{EP}), and finally, while most compounds are ionised at 70 eV, this is not always the case. 
Additionally, in an ideal case, the compounds should be ionised at a level representative of the ionisation dynamics where they would be used. In the case of \acs{HET}, although the electron energy is not constant, it can be considered 10\% of the voltage between electrodes \cite{Goebel2023}, being 30 eV a good reference value \cite{Goebel2008} for standard power. However, as just discussed, in the case of high-power \acs{HET}, the electron energy can be significantly higher with energies up to 100 eV being considered \cite{Goebel2008}. For \ac{GIE}, the electron energies tend to be lower from a few eV to 20 eV \cite{Goebel2023-2}.
So, while the ionisation energy used to train the algorithm is on the higher end of the usual ionisation energy, it is still within the range of \acs{HET} and of the same magnitude as \acs{GIE}. A more refined approach would involve generating a database with various excitation energies to train the model, enabling it to calculate the spectra at the relevant energy level.

However, with this imperfect dataset, it has still been possible to demonstrate the potential of \acs{ML} to predict the results of ionising a complex molecule in an electric thruster. In the future, a secondary layer of machine learning algorithms could also be implemented, combining the results from all predictions and classifying which chemical compounds are viable xenon substitutes. \cite{Giannetti2016}
Finally, one of the main issues of this section is the training database. The \acs{MS} NIST database, as the name states, is used to calibrate mass spectrometers, and in consequence, it is not necessarily the best to train the algorithms. Three of the main issues are that the spectra count the number of ions with a particular specific charge without reference to the number of ionised molecules, a similar issue is that there is no data on ionisation efficiency (especially relevant for \acs{EP}), and finally, while most compounds are ionised at 70 eV, this is not always the case. For the latter problem, a more refined approach would involve generating a database with various excitation energies to train the model.

However, with this imperfect dataset, it has still been possible to demonstrate the potential of \acs{ML} to predict the results of ionising a complex molecule in an electric thruster. In the future, a secondary layer of machine learning algorithms could be implemented, combining the results from all predictions and classifying which chemical compounds are viable xenon substitutes. \cite{Giannetti2016}

\section{Conclusions}

The project has implemented three different ML architectures, using the molecular fingerprint as the input, to predict the ionisation and fragmentation patterns of chemical compounds, particularly the ionisation energy, the minimum appearance energy, its associated ion mass, and the mass spectrum.

The ionisation energy was predicted with a mean relative error of 6.87\%, and 7.99\% for minimum appearance energy. Predictions derived from the ion mass associated with the minimum appearance energy exhibited a 23.89\% mean relative error.

Finally, the prediction of mass spectra showcased a cosine similarity of 0.6395, and the predicted outcome fell within the top 10 most similar mass spectra in 60.63\% of cases. Furthermore, focusing on compounds within a 30 Da range of the predicted mass, this recall@10 climbed to a notable 78\%.

To build upon insights gleaned from fragmentation pattern predictions, subsequent research could involve the implementation of a secondary layer of machine learning algorithms. These algorithms could be trained on combined results from all predictions, thus classifying chemical compounds as viable Xenon substitutes or not. A feasible method for classification, for instance, would involve employing trade-offs, as exemplified in \cite{Giannetti2016}.

In summary, the project showcased the predictive capabilities of ML to understand the behaviour of chemical compounds when ionised and presented three viable implementations, underscoring the potential of artificial intelligence to drive electric propulsion propellant development.

\section{Declarations}

This study was carried out without any external funding.

\section*{Nomenclature}

\begin{acronym}[Abbreviations]
\hypersetup{hidelinks}
\acro{AE}{appearance energy}
\acro{ECFPs}{extended circular fingerprints}
\acro{EI}{electron ionisation}
\acro{EP}{electric propulsion}
\acro{GIE}{Gridded Ion Engine}
\acro{HET}{Hall Effect Thruster}
\acro{LSTM}{long short-term memory}
\acro{ML}{machine learning}
\acro{MLP}{multi-layer perceptron}
\acro{MS}{mass spectrum}
\acro{IE}{Ionisation Energy}
\acro{NIST}{National Institute of Standards and Technology}
\acro{ReLU}{rectified linear unit}
\acro{SGD}{stochastic gradient descent}

\end{acronym}

\bibliography{sn-bibliography}

\end{document}